\documentclass[a4paper]{mem}
\usepackage{graphicx}
\usepackage[a4paper]{hyperref}

\begin{document}
   \title{HST multiband photometry of the globular cluster NGC 6388
}
   \author{G. Busso \inst{1}, 
   G.Piotto \inst{1},\\ 
          S. Cassisi \inst{2}\fnmsep
}

   \institute{Dipartimento di Astronomia, Universit\'a di Padova,
vicolo dell'Osservatorio 2, 35122 Padova; \email{busso@pd.astro.it, piotto@pd.astro.it} \\ 
         \and
            INAF - Osservatorio Astronomico di Collurania, via
M. Maggini, 64100 Teramo, Italy; \email{cassisi@te.astro.it}}

\abstract{We present color-magnitude diagrams (CMD) of the globular
  cluster NGC6388 based on HST multiband photometry (F255W, F335W,
  F439W, F555W). In this paper we focus our attention on the peculiar
  horizontal branch of this cluster.  After a careful reddening
  correction, we fitted the observed CMD with theoretical models. For the
  first time we demonstrated that the HB of this very metal rich globular
  cluster extends beyond $T_e>30.000$K, showing clear evidence of a 
  population of blue hook (\cite{D'Cruz}, \cite{Brown}) stars. Moreover, 
  we could demonstrate that the HB tilt (slope) is not a reddening effect, 
  and is present in all the photometric bands.

     }

  \authorrunning{G. Busso et al.}  
  \titlerunning{The globular cluster NGC6388} 
  \maketitle

\section{Introduction}

In this work we present preliminary results of an HST project 
(GO8718, PI Piotto) aimed at the investigation of the properties
of hot horizontal branch (HB) stars in a number of Galactic
globular clusters (GGC), based on HST WFPC2 multiband photometry.
Our first target was NGC~6388, a metal rich ([Fe/H]=-0.6) cluster 
with an anomalous HB. Despite its high metal content, this
GGC has an extended blue HB (\cite{Rich}), with a
significant tilt in the V vs. B-V bands (\cite{Raimondo}). 
The origin of both anomalies is not yet understood. In the
present paper, we will show that the HB tilt is visible in
all the photometric bands, and that it cannot be due to the strong reddening
affecting this cluster. We will also show that the HB of NGC~6388 extends well
beyond $T_e>30.000$K, with a broad tail, likely populated by
late-helium flashers.

 \section{Reduction and data analysis}

The new NGC~6388 data presented in this paper represents an extension
of the original F555W vs F439W-F555W color magnitude diagram (CMD)
published by \cite{GP}. In particular, within GO8718 we collected
ultraviolet images for a total exposure of 1060s in the F336W and of
9200s in the F255W bands.  Cosmic rays made difficult the analysis of
F255W data, in particular the identification of the stars and the PSF
profile computation. The data were reduced with DAOPHOT and ALLFRAMEII
(\cite{Stetson1}, \cite{Stetson2}). The PSF fitting photometry was then
calibrated to the WFPC2 flight photometric system following \cite{Dolphin}. 
The UV star list was finally cross-correlated with
the F439W and F555W list of stars, and only the stars identified
in all of the four bands have been used to construct the CMDs.
An example of the CMDs using different bands is shown in Fig.~1,
where different symbols show the same stars in the two panels.
The CMD from the UV bands shows the entire HB plus the brightest
blue stragglers (stars plotted as open circles). 

   \begin{figure}
   \centering      
   \includegraphics[width=7cm, bb=30 175 570 600,clip]{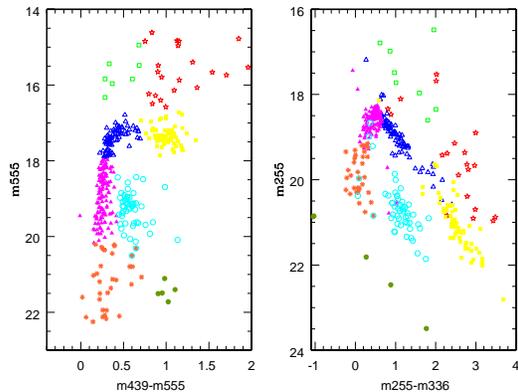}

      \caption{Optical and far-UV CMDs of the HB of NGC~6388}

    \end{figure}

Before comparing the observed CMDs with stellar evolutionary models, we need to apply
the appropriate reddening correction. As it is well known, the reddening correction depends
on the stellar temperature and the effect is more relevant for objects heavily
reddened, and larger for bluer photometric bands.
Therefore, we have determined the dependence of the reddening
correction on the temperature by convolving the appropriate atmospheric models of
\cite{Bessel98} with the filter bandpass, and applying the extinction
law of \cite{Scuderi96}. Fig.2 shows the size of the effect. For the
filters F555W and F439W there is almost no difference with a
constant reddening correction (we used the reddening values tabulated by
\cite{Holtzman95} for these bands); whereas for the bluer filters 
the difference is large, and becomes rather significant for the F255W
band.

After the reddening correction, we tried to match the observational data with stellar evolutionary models for HB structures. 
In the present work, we adopt the models computed for a metallicity Z=0.006 and an initial He content equal to Y=0.23 (for more details
on these models we refer to \cite{z99}). Bolometric magnitudes and effective
temperatures have been transformed into HST magnitudes according to the transformations 
provided by \cite{origlia}, based on the atmosphere models computed by
\cite{Bessel98}.
From \cite{Harris}, we adopted [Fe/H]=-0.6, a distance modulus $(m-M)_{F555W}$=16.54, and a reddening E(F439W-F555W)=0.37.

   \begin{figure} 
   \centering 
   \includegraphics[width=7cm]{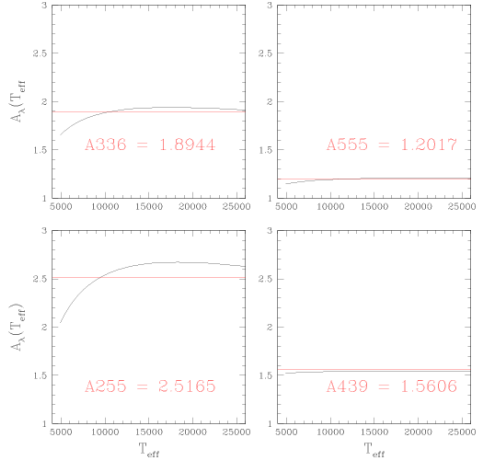}
   \caption{The four panels shows the extinction coefficients as a function
   of the temperature for the 4 photometric bands used in the present work.
   The constant line shows the average \cite{Holtzman95} coefficients for  
   the same bands.}
   \end{figure}

In Fig.~3 we show the comparison between the observed CMDs and the
models, adopting the quoted distance modulus.  Despite the fact that
we apply a temperature dependent absorption correction, it is not
possible to fit the entire HB.  In particular, while the models
properly reproduce the lower envelope of the red part of the HB, it is
not possible to fit the hot stars, which are always brighter than the
theoretical track. This is a different way to see the well known
problem of the tilted HB (\cite{Raimondo}): as shown by Fig. 3,
the HB presents a slope which cannot be reproduced by standard
models. This anomalous tilt of the HB is present in all the bands, and
cannot be an artifact of the reddening.  Also the differential
reddening can not account for it (see also the discussion in \cite{Raimondo}.  
If we try to fit the hottest HB
stars, the red HB would be fainter than predicted by the models.  A
slight HB slope is predicted by the canonical models
(\cite{Raimondo}), but it is much shallower than the observed
one. \cite{Sweigart} showed that the observed slope could be
reproduced assuming a large He abundance (Y=0.43). However, such an
high helium content is not consistent with other cluster observable, like
the position of the bump along the RGB (\cite{Riello}) and also with the
He abundance estimates based on the R parameter obtained by \cite{z99}.
A comparison between the UV CMD and the models shows that the
extended blue tail (EBT) of NGC~6388 reaches temperatures as
high as $T_e=31.500$ K (the hottest point in the models plotted
in Fig. 4).

   \begin{figure}
   \centering
   \includegraphics[width=5cm]{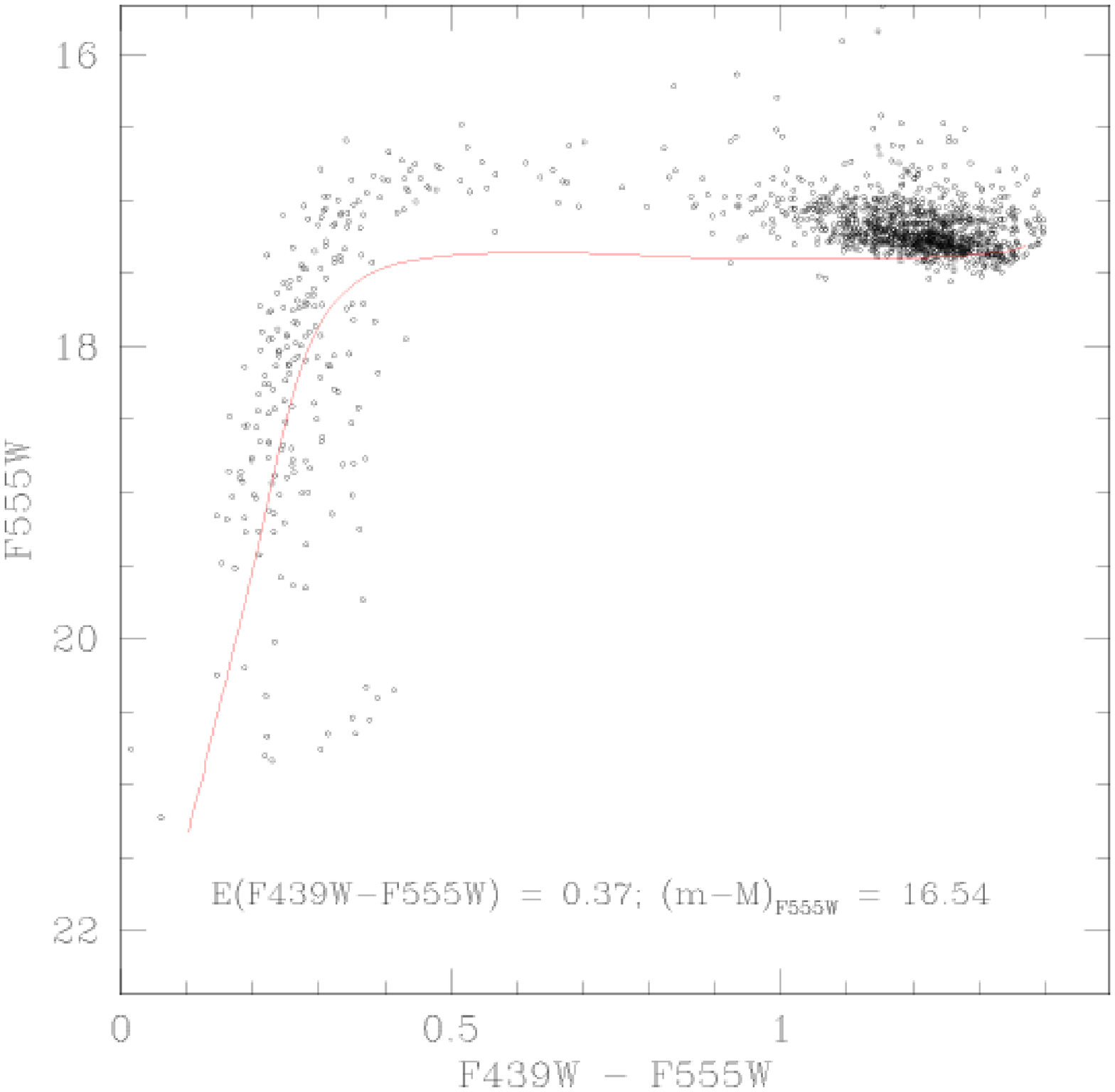}
   \includegraphics[width=5cm]{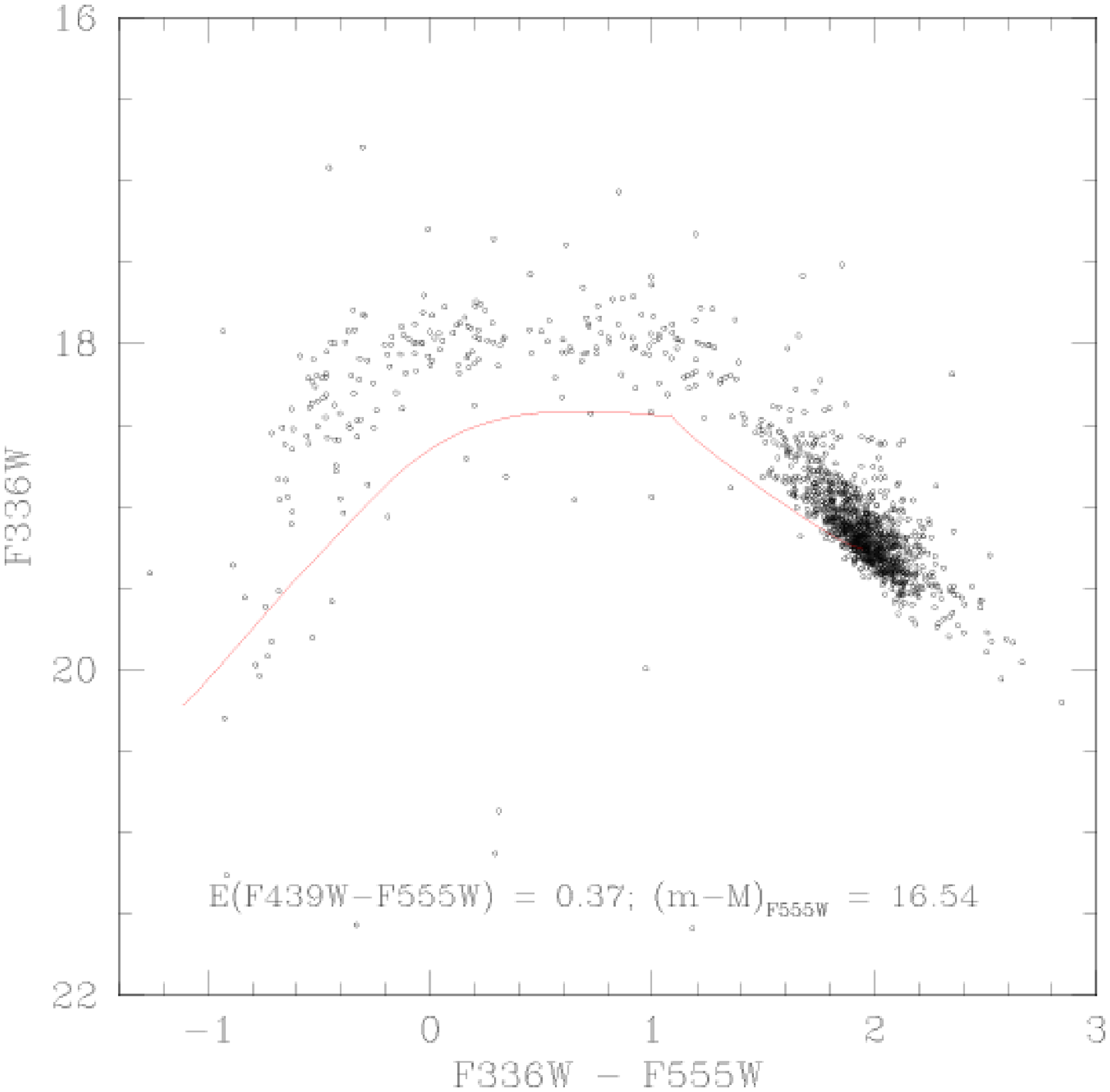}
   \includegraphics[width=5cm]{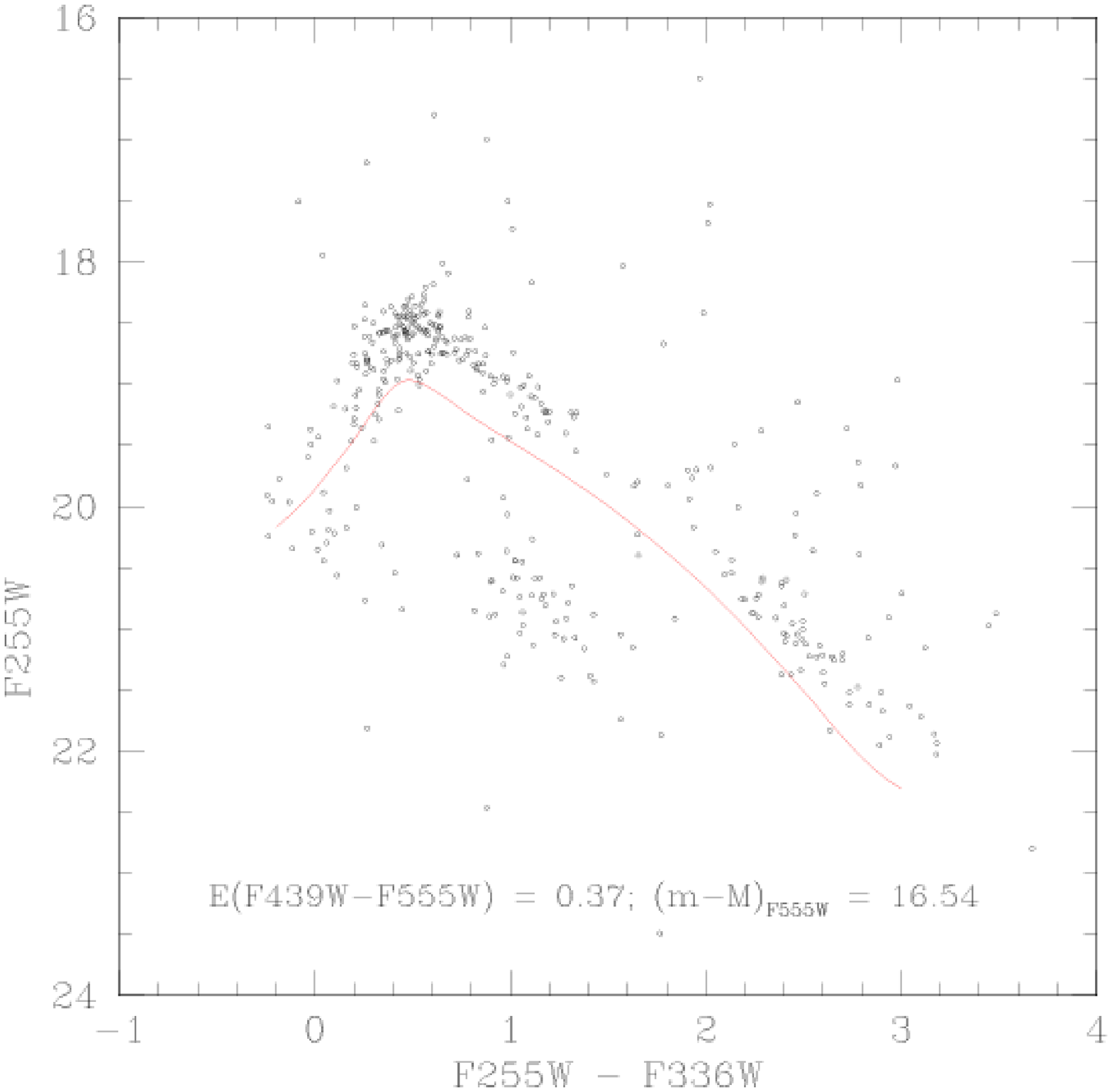}   
      \caption{Comparison of the observed CMDs with the models.}
    \end{figure}

The far-UV diagram shows another interesting feature. Even if we force
the models to fit the blue portion of the HB, the hottest stars appear
significantly fainter than predicted by the models,
and show a larger dispersion in temperature and/or luminosity.

A similar feature, called ``blue hook'',  has been
already observed in $\omega$ Cen (\cite{Moehler}) and NGC 2808
(\cite{Brown}). Brown et al. (2001) suggest that these stars in the
final part of the HB tail might be late helium-flashers. The late
helium flashers (\cite{D'Cruz} and \cite{Brown}) are stars 
that experience the He Flash while descending the white dwarf cooling 
sequence and then undergo He-mixing, and may have surface
abundance anomalies (\cite{C03}). This is the first time that late helium
flashers are identified in such a metal-rich GGC.

   \begin{figure} 
   \centering 
   \includegraphics[width=7cm]{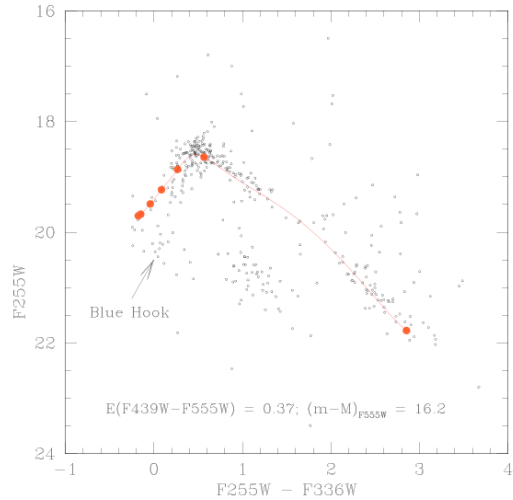}
   \caption{(F255W vs. F255W-F336W CMD: the full circles represent
   different temperatures, with a step of 5000 K, from $\sim$5000 to
   $\sim$31500 K; the arrow shows the {\itshape Blue Hook} candidate
   stars.} \end{figure}

\section{Conclusions}

The results of our analysis of HST multiband photometry of NGC 6388 can be
summarized as follows:
\begin{enumerate}
\item The HB tilt is not a reddening effect (at least not completely), and it is observed in all bands;
\item The Extended Blue Tail of this cluster reaches effective temperature greater than $T_e = 30000K$;
\item {\itshape Blue Hook} stars are observed also in a metal-rich 
GGC like NGC 6388.
\end{enumerate}

Preliminary results of NGC 6441, the globular cluster {\itshape twin} of NGC 6388 (they have very similar reddening and metallicity) show that the two clusters present analogous features (both present the tilt and the Blue Hook stars). A more detailed investigation and a thoughtful comparison with theoretical models is still in progress.

\bibliographystyle{aa}

\end{document}